\documentclass[11pt]{article}
\usepackage[utf8]{inputenc}

\usepackage[margin=1in]{geometry}

\usepackage{xcolor}
\title{Draft notes}
\date{}

\usepackage[english]{babel}
\usepackage{cite}
\usepackage{mcite}
\usepackage{graphicx}
\usepackage{amsmath,amssymb,amsfonts,mathrsfs}
\usepackage{shuffle}
\usepackage{ifdraft}

\allowdisplaybreaks[4]

\numberwithin{equation}{section}

\usepackage{amssymb}
\usepackage{amsmath}
\usepackage{bm}
\usepackage{color}
\usepackage{dcolumn}
\usepackage{graphics}
\usepackage{graphicx}
\usepackage{subfigure}
\usepackage{slashed}
\usepackage{placeins}
\usepackage{pdfpages}
\usepackage{eso-pic}
\usepackage{everyshi}
\usepackage{pdflscape}

\usepackage[v1compatible]{axodraw2}

\newcommand{\Eff}[5]{ 
    E_3  \left (\begin{smallmatrix}
   {#1} & {#3} \\
   {#2} & {#4}
\end{smallmatrix};#5\right)}

\newcommand{\ep}{\varepsilon}

\newcommand{\bea}{\begin{eqnarray}}
\newcommand{\eea}{\end{eqnarray}}
\newcommand{\be}{\begin{equation}}
\newcommand{\ee}{\end{equation}}

\begin{document}

\begin{center}
  {\Large {\bf
      On anomalous dimension
      %shortest four-point correlator
      in 3D ABJM model
      %the Checkerboard CFT
}}
\\ \vspace*{5mm} A.\ V.~Kotikov$^{1}$ and I.\ A.~Kotikov$^{2}$
%$^{a}$
\end{center}

\begin{center}
${}^{1}$
  Bogoliubov Laboratory of Theoretical Physics, Joint Institute for Nuclear Research, 141980 Dubna, Russia.\\
$^2$ Moscow State University, Faculty of Physics, 119991, Moscow, Russia
\end{center}

\begin{abstract}

% We compute perturbatively and numerically the
  Anomalous dimension of the shortest single-trace operator is
  calculated in a simple form for the fishnet limit of the twisted
ABJM model in 3D.
 
\end{abstract}

%\maketitle

\section{Introduction} %~~~~~ {\bf OLD}

Conformal field theories (CFTs) are widely used in modern high-energy
physics and statistical mechanics, where they describe many important
physical phenomena.
An extensive class of integrable theories in $d = 4$ was discovered in
\cite{Gurdogan:2015csr} as a special double scaling limit for $N = 4$ super Yang-Mills (SYM) theory
(weak coupling combined with strong $\gamma$-deformation). Later in
\cite{Kazakov:2018qbr}, the definition of these theories has been
extended to any $d$-dimensions.

In the planar ’t Hooft limit, the perturbation theory (PT) of such
CFTs is dominated by a very limited number of Feynman diagrams of a
certain shape: in the simplest of these theories, they are
represented by a regular square lattice.
As shown by A. Zamolodchikov \cite{Zamolodchikov:1980mb}, square lattice graphs are
integrable, since they are equivalent to statistical partition
functions in some integrable statistical-mechanical model of
continuous spin variables, which are the coordinates of the
vertices of the graph. These graphs are called fishnets because of
their shape, and the CFTs are called Fishnet CFTs.

In addition to Fishnet CFTs arising from the double scaling limit of the
$\gamma$-deformed  $N = 4$ SYM theory, an extensive class of so-called
Loom  Fishnet CFTs was proposed in \cite{Kazakov:2022dbd}. It is based on Zamolodchikov's
construction of integrable Feynman graphs of a more general type,
characterized by arbitrary vertex valence, any dimension $d$ and various types of propagators.

Recently in\cite{Alfimov:2023vev} , the Loom Fishnet CFTs class with
%M = 4 slopes, which have only
quartic scalar vertices, was studied. This model has been called the Checkerboard
CFT because all of its planar graphs are dual to the Baxter lattice
with only rectangular faces.

The theory has two coupling constants and four complex conjugate SU(N)
matrix fields $Z_j$ ($j=1,2,3,4$) with Lagrangian (\ref{LCB}).

The richness of the parameter space of the Checkerboard CFT allows
for various interesting reductions, which were discussed in
\cite{Alfimov:2023vev}.
So, for a specific choice of scaling dimensions  $\Delta_j$ of fields,
this theory reduces to a strongly twisted three-dimensional ABJM
Fishnet CFT
\cite{Caetano:2016ydc}, which will be studied below.

\section{Basics}

The Checkerboard Fishnet CFT studied in \cite{Alfimov:2023vev} is the theory of
four complex matrix scalar fields $Z_j$ of $N \times N$ components in
any space-time dimension $d$. The Lagrangian theory has general
non-local kinetic terms and two-quartic interactions
\be
L^{(CB)}=N {\rm Tr} \left[\sum_{j=1}^4
  \overline{Z}_j(-\partial_{\mu}\partial^{\mu})^{w_i}\,Z_j - \xi_1^2\,
  \overline{Z}_1 \overline{Z}_2Z_3Z_4 - \xi_2^2\,
  Z_1Z_2\overline{Z}_3 \overline{Z}_4\right]
\label{LCB}
\ee

The constraint $\omega_1+\omega_2+\omega_3+\omega_4=d$ is imposed
in order to work with dimensionless couplings $\xi_1^2$ and $\xi_2^2$.
Sometimes it is useful to switch from the parameters $\omega_j$ to
another set of labels (see \cite{Chicherin:2012yn}),
%, commonly used in the $SO(1, d + 1)$ spin-chain formalism [34],
namely
\be
\omega_1=u+d-\Delta_{+},~~\omega_2=-u+\Delta_{-},~~\omega_3=u+\Delta_{+},~~\omega_4-u-\Delta_{-}\,,
\label{Deltaj}
\ee
where $\Delta_{\pm}$ and the spectral parameter $u$ are generic. It follows from (\ref{LCB}) that the scaling
dimensions of the fields $\overline{Z}_j$ and $Z_j$ are $\Delta_j =d/2
-\omega_j$.

We are usually interested in the planar (or multicolored) limit of the
theory, $N \to \infty$.
For arbitrary $\omega_j$ values, the theory in (\ref{LCB}) is finite,
but there are special values when the correlators of the double trace
of length-2 operators diverge,
and a counter-term must be added to the Lagrangian in (\ref{LCB}).
This is the case when
\be
\Delta_{1}+\Delta_{2}=\frac{d}{2}=\Delta_{3}+\Delta_{4}~~~\mbox{or}~~~
\Delta_{1}+\Delta_{4}=\frac{d}{2}=\Delta_{2}+\Delta_{3}\,.
\label{d12} 
\ee

For example, if
%whenever
the first equation is satisfied, the double-trace  counter-terms
(see \cite{Alfimov:2023vev})
\be
L_{dt}^{(CB)}= \alpha(\xi_1,\xi_2)\,{\rm Tr}
\bigl(\overline{Z}_1\overline{Z}_2\bigr)\,{\rm Tr} \bigl(Z_3Z_4\bigr)
+\overline{\alpha}(\xi_1,\xi_2)\,\bigl(Z_1Z_2\bigr)\,{\rm Tr} \bigl(\overline{Z}_3\overline{Z}_4\bigr)
\label{LCBdt}
\ee
should be added to the Lagrangian (\ref{LCBdt}), and the couplings
$\alpha(\xi_1,\xi_2)$ adjusted to their critical values in order
to preserve conformal symmetry. The propagators of conjoint fields
$Z_k$ are
\be
D_i(x)=<Z_i(x)\overline{Z}_i(0)>=\frac{\Gamma(d/2-\omega_i)}{4^{\omega}\pi^{d/2}\Gamma(\omega_i)}\,\frac{1}{(x^2)^{d/2-\omega_i}}\,.
\label{Dix}
\ee

\subsection{The $L=2$ case}

Authors of \cite{Alfimov:2023vev} derived the exact expression for the shortest four-point correlator in the Checkerboard CFT, and extract the anomalous dimension of lightest
single-trace operator (using the methods of
\cite{Kazakov:2018qbr,Gromov:2018hut,Grabner:2017pgm})
%, which dominates the OPE s-channel,
\be
 {\rm Tr}\bigl[Z_1Z_2Z_1Z_2](x)\,.
\label{Z4} 
\ee
 
Within the choice of shortest length $L = 2$ and for operators of spin $S = 0$ the
solution $h(\nu,S=0)\equiv h(\nu)$ of the spectral
%function $h(\nu,S=0)\equiv h(\nu)$
problem reduces to computing a two-loop, massless, Kite master
integral with specific powers of the propagators.

Indeed, the eigenvalue $h(\nu)$
%of the so-called spectral equation function $h(\nu,S=0)\equiv h(\nu)$ 
 factor into the product of two terms
%(see appendix A for details),
\be
h(\nu)=h_1(\nu)\,h_2(\nu)\,,
\label{h12} 
\ee
where
\be
h_1(\nu)=B\left(\Delta_{1},\Delta_{2},\Delta_{3}+\Delta_{4}-\frac{\Delta}{2}\right),~~
h_2(\nu)=B\left(\Delta_{3},\Delta_{4},\Delta_{1}+\Delta_{2}-\frac{\Delta}{2}\right)\,
\label{h12_def} 
\ee
%where
and
\be
B(a_1,a_2,\tilde{\delta}) = \frac{\left(x_{00'}^2\right)^{\overline{\delta}+2a_1+2a_2-d}}{4^{d-2a_1-2a_2}(2\pi)^{2d}}\, \frac{1}{\bigl(a(a_1)a(a_2)\bigr)^2} \,
    \int \frac{d^dx_{1'}d^dx_{2'}}{(x_{1'2'}^2)^{\tilde{\delta}}(x_{01'}^2)^{a_{1}}(x_{1'0'}^2)^{a_{2}}(x_{0'2'}^2)^{a_{1}}(x_{2'0}^2)^{a_{2}}}\,,
\label{B_def} 
\ee
with
\be
a(a_i)=\frac{\Gamma(\tilde{a}_i)}{\Gamma(a_i)},~~(\tilde{a}_i=d/2-a_i)\,.
\label{ai} 
\ee

The result for anomalous dimension is obtained from the equation
%The spectral function obeys the following equation
\be
h(\nu)=\frac{1}{\zeta^4},~~\zeta^2=\zeta_1^2\zeta_2^2\,,
\label{hxi} 
\ee
which is in a sence a solution of full correlation function (see \cite{Alfimov:2023vev}).

\section{ABJM reduction}

As it was discussed in \cite{Alfimov:2023vev}, an interesting choice for parameters in (\ref{LCB}) is given by the formula
$\omega_4 = -u - \Delta_{-} = 0$.
In this case, the operators do not depend on $Z_4$, the path integral
over this field is purely Gaussian and can be exactly performed:
\be
\int DZ_4\,D\overline{Z}_4\, e^{\left(-\int {\rm
    Tr}[\overline{Z}_4\,Z_4-\xi_1^2\,\overline{Z}_1\,\overline{Z}_2\,Z_3\,Z_4
    -\xi_2^2\,Z_1\,Z_2\,\overline{Z}_3\,\overline{Z}_4]d^dx\right)}
\sim  e^{\left(-\int \xi_1^2\,\xi_2^2\,{\rm Tr}[\overline{Z}_3\,\overline{Z}_1\,\overline{Z}_2\,Z_3\,Z_1\,Z_2\,]d^dx\right)}\,,
\label{noZ4} 
\ee
where the proportionality constant does not depend on the fields and therefore 
irrelevant.

Then the Checkerboard CFT is staggered into the anisotropic
d-dimensional Fishnet CFT \cite{Kazakov:2022dbd}
\be
L_d^{(CB)} = N {\rm Tr} \bigl[\overline{Z}_1\,(-d_{\mu}d^{\mu})^{u+d-\Delta_{-}}\,Z_1+\overline{Z}_2\,(-d_{\mu}d^{\mu})^{-2u}\,Z_2+\overline{Z}_3\,(-d_{\mu}d^{\mu})^{u+\Delta_{+}}\,Z_3-\xi^2
  \overline{Z}_3\,\overline{Z}_1\,\overline{Z}_2\,Z_3\,Z_1\,Z_2\bigr]\,.
\label{LCB1} 
\ee
%where $\xi^2=\xi_1^2\,\xi_2^2$.
 Authors of \cite{Alfimov:2023vev} considered this theory as a
 d-dimensional generalisation of the 3d FCFT obtained  with the
 double-scaling in\cite{Caetano:2016ydc} from the ABJM theory,
 %with the double-scaling of large imaginary twist angle
%and weak coupling (up to replacing $Z_3$ with $\overline{Z}_3$). This
 %latter theory
 which is derived from
(\ref{LCB1}) at the point $d = 3$, $u = -1/2$, and $ \Delta_{+}= 3/2$ or, equivalently, $\Delta_{1}=\Delta_{2}=\Delta_{3} = 1/2$.

\subsection{ABJM $L = 2$ Fishnet}

%Here we explicitly represent
Let $h(\nu)=h_1(\nu)h_2(\nu)$,
where $h_1(\nu)$ and $h_2(\nu)$ are defined in Eqs. (\ref{h12_def})
and (\ref{B_def}).  According to (\ref{h12_def}), for the ABJM Fishnet
CFT
(see \cite{Caetano:2016ydc}), $h_1(\nu)$ and $h_2(\nu)$ are equal to \cite{Alfimov:2023vev}
\be
h_1(\nu)=B\left(\frac{1}{2},\frac{1}{2},2-\frac{\Delta}{2}\right),~~h_2(\nu)=B\left(\frac{1}{2},\frac{3}{2},1-\frac{\Delta}{2}\right),
\label{h12_L2} 
\ee
where $\Delta=3/2+2i\nu$ for $d=3$. The function $B(a_1,a_2,\tilde{\delta})$ is defined in Eqs. (\ref{B_def}) and (\ref{ai}).

As shown in \cite{Alfimov:2023vev},
calculating $h_2$ is trivial in this particular case. Indeed, starting with
the integral representation
(\ref{B_def}) and using the following representation for the  Dirac $\delta$-function,
\be
%\varlim_{a_2 \to d/2}
%\varinjlim_{a_2 \to d/2}
\lim_{a_2 \to d/2}
\frac{\Gamma(a_2)}{\Gamma(d/2-a_2)x^{2a_2}}=\pi^{d/2} \delta^{(d)}(x)\,,
\label{lim} 
\ee
one can easily see that  \cite{Alfimov:2023vev},
\be 
h_2(\nu)=\frac{1}{16\pi^2}\,.
\label{h2_L2} 
\ee

We now turn our attention to $h_1(\nu)$. In Ref.
\cite{Alfimov:2023vev} following the results in \cite{Derkachev:2022lay},
$h_1(\nu)$ was represented as a combination
of two-fold series. Below we will obtain a one-fold-series
representation, based on the results in  \cite{Kotikov:1995cw}.

%{\bf END OLD}

\section{Results for $h_1(\nu)$}

To find $h_1$ it is necessary to calculate the Feynman integral
$J(\tilde{\delta},x^2)$, ($\tilde{\delta}=2-\Delta/2$),
%=5/4-i\nu$),
which is as follows

\vskip 0.5cm
\be
J(\tilde{\delta},x^2) \, =
 \hspace{3mm}
%\frac{2^{n-m} \Gamma(n-m+\alpha)}{(n-m)!\Gamma(\alpha)} \, \hspace{3mm}
\raisebox{1mm}{{
    %\begin{picture}(90,30)(0,4)
    \begin{axopicture}(90,10)(0,4)
%  \SetWidth{2.0}
  \SetWidth{0.5}
%\CArc(5,5)(80,20,160)
%\CArc(45,5)(40,0,180)
%\Arc(45,-7)(40,20,160)
\Arc(45,-7)(40,20,90)
%\Arc[arrow](45,-7)(40,90,160)
\Arc(45,-7)(40,90,160)
 \SetWidth{1.5}
\Line(45,-25)(45,35)
 \SetWidth{0.5}
%\Line[arrow](40,5)(85,5)
\Arc(45,17)(40,200,270)
%\Arc[arrow](45,17)(40,270,340)
\Arc(45,17)(40,270,340)
%\Arc(45,17)(40,200,340)
\Vertex(45,-23){2}
\Vertex(45,33){2}
\Vertex(5,5){2}
\Vertex(85,5){2}
%\SetWidth{1.0}
%\Vertex(5,5){2}
%\Line(5,5)(-5,5)
%\Line(85,5)(95,5)
%Line(45,-25)(45,-35)
%Line(85,5)(95,10)
%\Vertex(40,5){2}
%\Vertex(40,15){2}
%\Vertex(40,-15){2}
\Text(15,25)[b]{$\scriptstyle 1/2$}
\Text(15,-25)[b]{$\scriptstyle 1/2$}
%\Text(55,5)[b]{$\scriptstyle \tilde{\delta}$}
\Text(55,5)[b]{$\tilde{\delta}$}
\Text(75,25)[b]{$\scriptstyle 1/2$}
\Text(75,-25)[b]{$\scriptstyle 1/2$}
%\Text(75,-22)[t]{$\alpha$}
%\Text(75,-22)[t]{$\scriptstyle \alpha$}
\Text(87,-10)[b]{$x$}
\Text(0,-10)[b]{0}
%\Text(93,-5)[b]{$\to$}
%\Text(93,-12)[b]{$q$}
%\Text(93,20)[b]{$\to$}
%\Text(93,12)[b]{$p$}
%\Text(52,-30)[b]{$\to$}
%\Text(52,-38)[b]{$p$}
%\Text(-3,-5)[b]{$p$}
\end{axopicture}
}}~~= \frac{C_{J}(\tilde{\delta})}{x^{2\overline{\delta}}},~~\overline{\delta}=2+\tilde{\delta}-d=\tilde{\delta}-1 \, ,
\label{Jtdelta}
\ee
\vskip 1cm
%,
\noindent
where
\be
\delta=\frac{1}{4}+i\nu,~~\tilde{\delta}=\frac{d}{2}-\delta=\frac{5}{4}-i\nu
\label{delta}
\ee

Fourier transform relates the diagram $J(\tilde{\delta},x^2)$ with the one $I(\delta,p^2)$:
\vskip 0.5cm
\be
I(\delta,p^2) \, =
 \hspace{3mm}
%\frac{2^{n-m} \Gamma(n-m+\alpha)}{(n-m)!\Gamma(\alpha)} \, \hspace{3mm}
\raisebox{1mm}{{
    %\begin{picture}(90,30)(0,4)
    \begin{axopicture}(90,10)(0,4)
%  \SetWidth{2.0}
  \SetWidth{0.5}
%\CArc(5,5)(80,20,160)
%\CArc(45,5)(40,0,180)
%\Arc(45,-7)(40,20,160)
\Arc(45,-7)(40,20,90)
%\Arc[arrow](45,-7)(40,90,160)
\Arc(45,-7)(40,90,160)
 \SetWidth{1.5}
\Line(45,-25)(45,35)
 \SetWidth{0.5}
%\Line[arrow](40,5)(85,5)
\Arc(45,17)(40,200,270)
%\Arc[arrow](45,17)(40,270,340)
\Arc(45,17)(40,270,340)
%\Arc(45,17)(40,200,340)
\Vertex(45,-23){2}
\Vertex(45,33){2}
\Vertex(5,5){2}
\Vertex(85,5){2}
%\SetWidth{1.0}
%\Vertex(5,5){2}
\Line(5,5)(-5,5)
\Line(85,5)(95,5)
%Line(45,-25)(45,-35)
%Line(85,5)(95,10)
%\Vertex(40,5){2}
%\Vertex(40,15){2}
%\Vertex(40,-15){2}
%\Text(15,25)[b]{$\scriptstyle 1/2$}
%\Text(15,-25)[b]{$\scriptstyle 1/2$}
\Text(55,5)[b]{$\delta$}
%\Text(75,25)[b]{$\scriptstyle 1/2$}
%\Text(75,-25)[b]{$\scriptstyle 1/2$}
%\Text(75,-22)[t]{$\alpha$}
%\Text(75,-22)[t]{$\scriptstyle \alpha$}
%\Text(87,-10)[b]{$x$}
%\Text(0,-10)[b]{0}
\Text(-3,-5)[b]{$\to$}
\Text(-3,-12)[b]{$q$}
%\Text(93,-5)[b]{$\to$}
%\Text(93,-12)[b]{$q$}
%\Text(93,20)[b]{$\to$}
%\Text(93,12)[b]{$p$}
%\Text(52,-30)[b]{$\to$}
%\Text(52,-38)[b]{$p$}
%\Text(-3,-5)[b]{$p$}
\end{axopicture}
}}~~= \frac{C_{I}(\delta)}{p^{2(\delta+1)}},~~
%\overline{\delta}=2+\tilde{\delta}-d=\tilde{\delta}-1 \, ,
\label{Idelta}
\ee
\vskip 1cm
%,
\noindent
where hereafter (see, for example, \cite{Kotikov:2018wxe})
\be
C_{J}(\tilde{\delta})\stackrel{F}{=} K(\tilde{\delta})\,C_{I}(\delta)
\label{Ctdelta.1}
\ee
and
\be
 K(\tilde{\delta})=
\frac{a(\tilde{\delta})a(1/2)}{a(\tilde{\delta}-1)}\,=\frac{1}{\pi^2\delta(1/2-\delta)}\,,~~a(\delta)=\frac{\Gamma(\tilde{\delta})}{\Gamma(\delta)},~~\tilde{\delta}=\frac{d}{2}-\delta\,.
\label{Ctdelta.2}
\ee
In (\ref{Ctdelta.1}) we have shown exactly that this relation was
obtained using the Fourier transform.

We note that it is common practice in CFT to work in $x$-space.
%We note that we want to work in $x$-space.
To do it, we
can consider the diagram  $I(\delta,x^2)$ in $x$-space, which is dual
to $I(\delta,p^2)$, and which can be obtained
using so-called dual transform \cite{Kotikov:1995cw,Kazakov:1986mu}, where all momenta
are replaced by coordinates.
\footnote{It is now more popular to use a dual transform as
$p_i=(x_i-x_{i+1})$ (see, for example, \cite{Drummond:2008vq}).}
So, %we have
the dual transform does not change the Feynman integrals but it can change their graphic representations.
In the case of  $I(\delta,x^2)$, the graphical representation
is not changed and we have
\vskip 0.5cm
\be
I(\delta,x^2) \, =
%\overset{D}{=}
\hspace{3mm}
%\frac{2^{n-m} \Gamma(n-m+\alpha)}{(n-m)!\Gamma(\alpha)} \, \hspace{3mm}
\raisebox{1mm}{{
    %\begin{picture}(90,30)(0,4)
    \begin{axopicture}(90,10)(0,4)
%  \SetWidth{2.0}
  \SetWidth{0.5}
%\CArc(5,5)(80,20,160)
%\CArc(45,5)(40,0,180)
%\Arc(45,-7)(40,20,160)
\Arc(45,-7)(40,20,90)
%\Arc[arrow](45,-7)(40,90,160)
\Arc(45,-7)(40,90,160)
 \SetWidth{1.5}
\Line(45,-25)(45,35)
 \SetWidth{0.5}
%\Line[arrow](40,5)(85,5)
\Arc(45,17)(40,200,270)
%\Arc[arrow](45,17)(40,270,340)
\Arc(45,17)(40,270,340)
%\Arc(45,17)(40,200,340)
\Vertex(45,-23){2}
\Vertex(45,33){2}
\Vertex(5,5){2}
\Vertex(85,5){2}
%\SetWidth{1.0}
%\Vertex(5,5){2}
%\Line(5,5)(-5,5)
%\Line(85,5)(95,5)
%Line(45,-25)(45,-35)
%Line(85,5)(95,10)
%\Vertex(40,5){2}
%\Vertex(40,15){2}
%\Vertex(40,-15){2}
%\Text(15,25)[b]{$\scriptstyle 1/2$}
%\Text(15,-25)[b]{$\scriptstyle 1/2$}
\Text(55,5)[b]{$ \delta$}
%\Text(75,25)[b]{$\scriptstyle 1/2$}
%\Text(75,-25)[b]{$\scriptstyle 1/2$}
%\Text(75,-22)[t]{$\alpha$}
%\Text(75,-22)[t]{$\scriptstyle \alpha$}
\Text(87,-10)[b]{$x$}
\Text(0,-10)[b]{0}
%\Text(-3,-5)[b]{$\to$}
%\Text(-3,-12)[b]{$q$}
%\Text(93,-5)[b]{$\to$}
%\Text(93,-12)[b]{$q$}
%\Text(93,20)[b]{$\to$}
%\Text(93,12)[b]{$p$}
%\Text(52,-30)[b]{$\to$}
%\Text(52,-38)[b]{$p$}
%\Text(-3,-5)[b]{$p$}
\end{axopicture}
}}~~= \frac{C_{I}(\delta)}{x^{2(\delta-1)}}\,.~~
%\overline{\delta}=2+\tilde{\delta}-d=\tilde{\delta}-1 \, ,
\label{Idelta.du}
\ee
\vskip 1cm
%,
\noindent
%because after dual transform the diagram $I(\delta)$ does not change.

The result for $C_{I}(\delta)$ has the following form
\cite{Kotikov:1995cw} (see also a review in
\cite{Grozin:2012xi})
\be
(4\pi)^d C_{I}(\delta)= 2\pi
%\Gamma(\delta)
\,\left[\frac{\sqrt{\pi}\Gamma(1-\delta)\Gamma^2(\delta)}{\Gamma(1/2+\delta)}- S(\delta)\right]\,,
  \label{CIdelta}
\ee
where
\be
S(\delta)=\Gamma(\delta)\,
\sum_{n=0}^{\infty} \,
  \frac{\Gamma(n+1)}{\Gamma(n+\delta+1)}\, \frac{1}{(n+\delta+1/2)}\,,
  \label{Jtdelta}
\ee
which can be represented as ${}_3F_2$-hypergeometric function with argument 1.
\footnote{The results for $I(\delta,x^2)$ had been firstly obtained \cite{Kazakov:1983pk} in a form, including
  two ${}_3F_2$-hypergeometric  functions with argument $-1$. An exact coincidence of the result and the one shown in
  (\ref{CIdelta}) has been proven in Ref. \cite{Kotikov:2018uat}.}
  
Evaluating $S(\delta)$ (see Appendix A) we have
\be
%(32\pi^2)
C_{I}(\delta)= \frac{1}{(32\pi^2)}\,
 \int_1^{\infty}\,dz\,\left[z^{\delta-1}+z^{-1/2-\delta}\right]\,\frac{1}{\sqrt{1+z}}\,
\ln \frac{\sqrt{1+z}+1}{\sqrt{1+z}-1}\,,
  \label{CIdelta.1}
\ee
where the property $C_{I}(\delta)=C_{I}(1/2-\delta)$ is exactly shown.
\footnote{The property  $C_{I}(\delta)=C_{I}(1/2-\delta)$ can be  exactly found 
in Eq. (\ref{CIdelta})
 using properties of
  ${}_3F_2$-hypergeometric function with argument 1 (see \cite{Prudnikov}).}
  %  A derivation of this relation using transformations of the corresponding Feynman integrals can be found in Appendix C.
%}
The property together with the result (\ref{Ctdelta.2}) for the coefficient $K(\tilde{\delta})$ leads to the one
$C_{J}(\tilde{\delta})=C_{J}(5/2-\tilde{\delta})$ since the replacement $\delta\leftrightarrow 1/2-\delta$ corresponds to the one
$\tilde{\delta}\leftrightarrow 5/2-\tilde{\delta}$.

The result for $h_1(\nu)$ has the form
\be
h_1(\delta)=\frac{\pi^2}{4} C_{J}(\tilde{\delta})=\frac{1}{2\delta(1-2\delta)} C_{I}(\delta)\,,
  \label{h1delta}
\ee
where $C_{I}(\delta)$ is given above in Eq. (\ref{CIdelta.1}).

Following \cite{Alfimov:2023vev} we consider the variables $\zeta(\delta)$ and $\xi(\delta)$ as
%So, for
$\zeta(\delta)=1/h(\delta)$ (see Eq. (\ref{hxi}))
with $h(\delta)=h_1(\delta)h_2(\delta)=h_1(\delta)/(16\pi^2)$ and
$\xi(\delta)=\zeta(\delta)/(256\pi^2)$. So, we have
\be
\zeta(\delta)=\frac{32\pi^2\delta(1-2\delta)}{C_{I}(\delta)}\,,~~
% \label{zeta_delta}
%\ee
%So, we obtain
%\be
\xi(\delta)=\frac{\zeta(\delta)}{256\pi^2}=\frac{\delta(1-2\delta)}{8C_{I}(\delta)}\,.
 \label{xi_delta_ge}
\ee

The result for $\xi(\delta)$ is shown in Fig. 1 as a solid line. The
inverse function $\delta(\xi)$ is shown in Fig. 2 below. It defines
the values of the anomalous dimension $\gamma$ ($\Delta=2+\gamma$). So,
we have that $\gamma=\Delta-2=2\delta-1$ and thus we get
$\gamma(\xi)$ from Eq. (\ref{xi_delta_ge}).

\begin{figure}[!t]
%\begin{figure}[!htb]%[t]%[tl]
%    \includegraphics[width=0.18\textwidth]{fig1}
\centering
\includegraphics[width=0.58\textwidth]{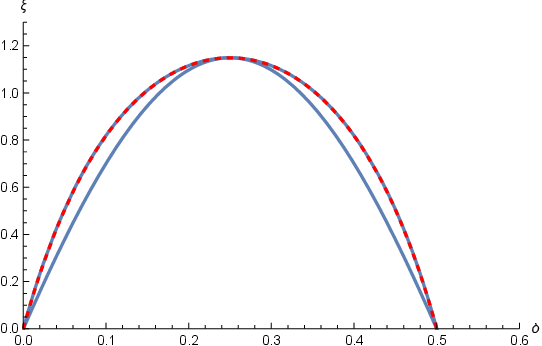}
\caption{\label{fig:C}
  The results for $\xi(\delta)$ with the exact values (\ref{CIdelta.1}) for $C_{I}(\delta)$ (solid blue line)
  and with the first two and three terms in the expansion (\ref{CIdelta.3}) (dotted read and blue lines, respectively).  }
%  \end{center}
\end{figure}

\subsection{The case $\delta=1/4+\ep$}

Now consider $C_{I}(\delta)$ in a vicinity of the critical point $\delta_{c}=1/4$.
Taking $\delta=1/4+\ep$, we have from (\ref{CIdelta.1})
\bea
(32\pi^2) C_{I}(\delta)&=&
 \int_1^{\infty}\,dz\,\left[z^{\ep}+z^{-\ep}\right]\,\frac{z^{-3/4}}{\sqrt{1+z}}\,
 \ln \frac{\sqrt{1+z}+1}{\sqrt{1+z}-1}\nonumber \\
 &=&2 \int_1^{\infty}\,dz\,\frac{z^{-3/4}}{\sqrt{1+z}}\,
\ln \frac{\sqrt{1+z}+1}{\sqrt{1+z}-1}\, \sum_{m=0}^{\infty} \frac{\ep^{2m}}{(2m)!} \ln^{2m}z,.
  \label{CIdelta.2}
\eea

It is conveninet to change the integration variable as
\be
z \to t=\frac{\sqrt{1+z}-1}{\sqrt{1+z}+1}, ~~\mbox{and, so,}~~ \sqrt{1+z}=\frac{1+t}{1-t},~~z=\frac{4t}{(1-t)^2},~~\frac{dz}{dt}=4\frac{1+t}{(1-t)^3}\,,
 \label{z_t}
\ee
that gives
\be
C_{I}(1/4+\ep)= \sum_{m=0}^{\infty} \frac{\ep^{2m}}{(2m)!} \, C^{(m)}_{I}(1/4)\,,
\label{CIdelta.3}
\ee
where
\bea
%(32\pi^2)
&&C^{(m)}_{I}(1/4)= 
-\frac{\sqrt{2}}{(16\pi^2)}\, \int_a^{1}\frac{dt\,\ln t}{t^{3/4}\sqrt{1-t}}\,\ln^{2m}\left[\frac{4t}{(1-t)^2}\right]\nonumber \\
&&=-\frac{\sqrt{2}}{(4\pi^2)}\, \int_{\sqrt{a}}^{1}\frac{dp\,\ln p}{\sqrt{p(1-p^2)}}\, \ln^{2m}\left[\frac{4p^2}{(1-p^2)^2}\right]
%\nonumber \\&=&
=- \frac{\sqrt{2}}{\pi^2}\,\int_{a^{1/4}}^{1}\frac{ds\,\ln s}{\sqrt{1-s^4}}\, \ln^{2m}\left[\frac{4s^4}{(1-s^4)^2}\right]\,,
\label{CIdelta.3a}
\eea
where we used the following variables:
\be
t=p^2=s^4,~~a=\frac{\sqrt{2}-1}{\sqrt{2}+1}\,.
\label{p_t_a}
\ee
Notice that rather similar results have been obtained for on-shall
massive Feynman integrals in the framework of Nonrelativistic QCD in \cite{Kniehl:2005bc}.
The results (\ref{CIdelta.3a}) can be represented in the form of
Elliptic Polylogarithms (see Ref. \cite{Bourjaily:2022bwx} and
references and discussions therein), as it will be shown in the
following subsection, as well as in terms of itrated integrals with
algebraic kernels \cite{Bezuglov:2020ywm}.

Using the first two and three terms of the expansion (\ref{CIdelta.3})
we can approximate $C_{I}(1/4+\ep)$ and, using it, to obtain the
%approximate
values for $\xi(\delta)$ in the following form
\be
\xi(\delta)=\frac{(1-16\ep^2)}{64C_{I}(1/4+\ep)}\,.
 \label{xi_delta_ap}
\ee

The results for $\xi(\delta)$
with the first two and three terms in the expansion (\ref{CIdelta.3}) are presented in Fig.1 and shown by dotted read and blue lines, respectively.
We see coincidence between the exact result for $\xi(\delta)$ and approximations in the points $\delta=0$, $1/4$ and $1/2$. This coincidence in the points
$\delta=0$ and $1/2$ has the place since we did not expand the numerator in Eq. (\ref{xi_delta_ap}).

\subsection{The case $\delta=\delta_c=1/4$}

At the critical point $\delta_{c}=1/4$ we have
\be
C_{I}(1/4)\equiv C^{(0)}_{I}(1/4)\approx 0.013599~~\mbox{ and, so, }~~
\zeta(\delta_c)\approx 2903,~~ \xi(\delta_c)\approx 1.149\,,
 \label{CI14o}
\ee
that is in full agreement with ones obtained in Ref. \cite{Alfimov:2023vev}.

Moreover, we note that the result for $C_{I}(1/4)$ can be represented in the following form
\be
 C_{I}(1/4)=-\frac{\sqrt{2}}{4\pi^2}\, \int_{\sqrt{a}}^{1}\frac{dp\,\ln p}{\sqrt{p(1-p^2)}}\,=-\frac{\sqrt{2}}{4\pi^2}\,\Bigl[
   \Eff{0}{0}{0}{1}{1}-\Eff{0}{0}{0}{1}{\sqrt{a}}\Bigr]\,,
 \label{CI14}
\ee
where $\Eff{0}{0}{0}{1}{\sqrt{a}}$ are Elliptic Dilogarithm. The terms $C^{(m)}_{I}(1/4)$ $m\geq 1$ in r.h.s. of
(\ref{CIdelta.3})
can be also represented as
multiple Elliptic Polylogarithms \cite{Broedel:2017kkb}, using a
shuffle algebra for
subinttegral expressions (see, for example, \cite{Campert:2020yur}).

We also note that the term $\Eff{0}{0}{0}{1}{1}$ can be simplified to the form
\be
\Eff{0}{0}{0}{1}{1}\equiv \int_{0}^{1}\frac{dp\,\ln p}{\sqrt{p(1-p^2)}}=\frac{2\sqrt{\pi}\Gamma(1/4)}{\Gamma(3/4)}\left[\Psi\left(\frac{1}{4}\right)-\Psi\left(\frac{3}{4}\right)\right]
=-2\sqrt{\pi}\,\Gamma^2(1/4)\,,
 \label{E31}
\ee
where we used the properties of the Riemann's $\Gamma$- and $Psi$-functions:
\be
\Gamma(2z)=\frac{2^{2z-1}}{\sqrt{\pi}}\, \Gamma(z)\,\Gamma(z+1/2),~~\Psi(1-z)-\Psi(z)=\pi Cot(\pi z)\,,
 \label{Gamma2z}
 \ee
 for the last step of evaluations.

\subsection{Small $\delta$ values}

The authors of \cite{Alfimov:2023vev} considered the small
$\gamma$-expansion of $h_1(\nu)$.
Since $h_1(\nu)$ is symmetric for $\gamma \leftrightarrow 1+ \gamma$ and
then $1+\gamma=2\delta$.
%, then $1+ \gamma=2\delta$.
Therefore, it is enough to consider simply the $\delta$-expansion.
The $\gamma$-expansion will be shown in the following subsection.

It is convenient to calculate $C_I(1/2-\delta)$ since $C_I(1/2-\delta)=C_I(\delta)$, i.e.
\be
(32\pi^2) C_{I}(1/2-\delta)=
%2\pi
%\Gamma(\delta)\,\left[
\frac{\sqrt{\pi}\Gamma(1/2+\delta)\Gamma^2(1/2-\delta)}{\Gamma(1-\delta)}- \overline{S}(\delta)
%\right]
\,,
  \label{CI12delta}
\ee
where
\be
\overline{S}(\delta)=S(1/2-\delta)=\Gamma(1/2-\delta)\,
\sum_{n=0}^{\infty} \,
  \frac{\Gamma(n+1)}{\Gamma(n+3/2-\delta)}\, \frac{1}{(n-\delta+1)}
  \label{oSdelta}
\ee

Using Eq. (\ref{Gamma2z}) for $\Gamma(2z)$,
we have for the first term in the r.h.s. of (\ref{CI12delta})
\be
\frac{\sqrt{\pi}\Gamma(1/2+\delta)\Gamma^2(1/2-\delta)}{\Gamma(1-\delta)}=\pi^2\,2^{2\delta}\,\frac{\Gamma^2(1-2\delta)\Gamma(1+2\delta)}{\Gamma^3(1-\delta)\Gamma(1+\delta)}\,.
 \label{FTerm}
 \ee

 To expand in $\delta$ we use the following property:
 \be
 \Gamma(1+\delta)=\exp\left[-\gamma_{\rm E} \delta+\sum_{m=2}^{\infty} \, \frac{\zeta_m}{m}\, (-\delta)^m\right]\,,
 \label{Gamma1d}
\ee
where $\gamma_{\rm E}$ and $\zeta_m$ are the Euler constant and Euler $\zeta$-functions.

Thus, we have for eq. (\ref{FTerm})
\be
\frac{\sqrt{\pi}\Gamma(1/2+\delta)\Gamma^2(1/2-\delta)}{\Gamma(1-\delta)}=\pi^2\,2^{2\delta}\, \exp\left[\sum_{m=2}^{\infty} \, \zeta_m\,p_m \delta^m\right]
=\pi^2\,2^{2\delta}\, \exp\left[2l_2\delta+\sum_{m=2}^{\infty} \, \zeta_m\,p_m \delta^m\right]
\,,
 \label{FTerm.1}
 \ee
 where
 \be
l_2=\ln 2,~
p_m=\frac{1}{m}\Bigl[2^{m+1}-3+(-1)^m\bigl(2^m-1\bigr)\Bigr],~~\mbox{and, so,}~~
p_2=4,~p_3=2,~p_4=11,~p_5=6\,.
 \label{pm}
 \ee

 The results for  $\overline{S}(\delta)$ have the following form (see Appendix B):
 \be
 \overline{S}(\delta)
 %&=& \sum_{s=0}^{\infty}\, \delta^s \, \int_0^1\,dt\,\frac{t^{-1/2-\delta}}{1-t}\,{\rm Li}_{s+1}(1-t)
 =
 %2 \sum_{s=0}^{\infty}\, \delta^s \, \int_0^1\,dp\,\frac{p^{-2\delta}}{1-p^2}\,{\rm Li}_{s+1}(1-p^2)\nonumber \\&=&
 2 \sum_{s=0}^{\infty}\, \delta^s \, \int_0^1\,\frac{dp}{1-p^2}\,\sum_{m=0}^{\infty}\,\frac{(-2)^m}{m!}\,\ln^mp\,{\rm Li}_{s-m+1}(1-p^2)\,.
\label{oSdelta.C3f}
\ee

From (\ref{FTerm.1}) and (\ref{oSdelta.C3f}) we see that for each $\delta$-degree there are terms with same level
of transcendentality (see, for example, \cite{Kotikov:2000pm} for most popular usage of the property, Ref.
\cite{Fleischer:1998nb} for application for Feynman diagrams and \cite{Kotikov:2023uhr} for a recent review). 

\begin{figure}[!t]%[t]%[tl]
%\begin{figure}[!htb]%[t]%[tl]
%    \includegraphics[width=0.18\textwidth]{fig1}
\centering
\includegraphics[width=0.58\textwidth]{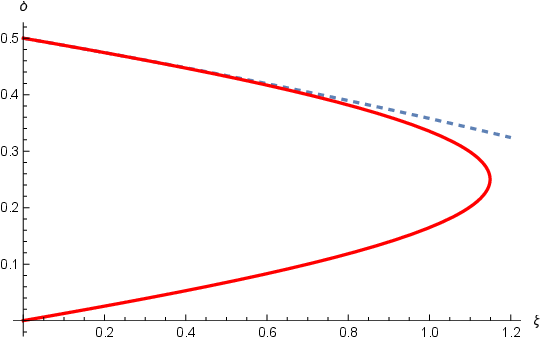}
\caption{\label{fig:B}  The results for $\delta(\xi)$ based on  (\ref{xi_delta_ge}) with the exact result (\ref{CIdelta.1}) for
$C_{I}(\delta)$ (solid line) and the approximation (\ref{xi}) (dashed line).
    }
%  \end{center}
\end{figure}

Taking three terms in expansions (\ref{FTerm.1}) and (\ref{oSdelta.C3f}) we have
\be
C_I(\delta)=\frac{1}{64\pi^2} \bigl[C_0+2\delta\, C_1 + 4\delta^2\, C_2 + O(\delta^3)\bigr]\,,
\label{CIdelta.O3}
\ee
where
%($l_2=\ln 2$)
\be
C_0=\pi^2,~~C_1=\pi^2\,l_2-\frac{21}{2}\,\zeta_3,~~C_2=\frac{\pi^4}{40}+\,\frac{l_2^4}{2}+12{\rm Li}_4(1/2),~~(l_2\equiv \ln 2)\,,
\label{Ci}
\ee
where ${\rm Li}_k$ are Polylogarithms.

Using the results, we see that
\be
\xi(\delta)
%\equiv \frac{\zeta(\delta)}{256\pi^2}
\approx \frac{8\pi^2\delta(1-2\delta)}{\bigl[C_0+2\delta\, C_1 + 4\delta^2\, C_2\,
    %+ O(\delta^3)
    \bigr]}
\label{xi}
\ee
which is an appromimation of the exact results in (\ref{xi_delta_ge})
and (\ref{xi_delta_ap}).
%\be
%\xi(\delta)=\frac{\delta(1-2\delta)}{8C_{I}(\delta)}\,,
% \label{xi_delta}
%\ee
%obtained from (\ref{zeta_delta}) with $C_{I}(\delta)$ from (\ref{CIdelta.1}).

In Fig.2, we show the inverse dependence
$\delta(\xi)$ using Eq. (\ref{xi_delta_ge}) with the exact result (\ref{CIdelta.1}) for
$C_{I}(\delta)$ and the approximate value (\ref{xi}). We see that
the  approximation is quite accurate, except for the vinicity of the ctitical point
($\delta=\delta_c=1/4,\xi=\xi_c$). We note that our results are
completely consistent with results in Ref. \cite{Alfimov:2023vev}
after replacing $\delta \to \gamma=2\delta-1$.

\subsection{Small $\gamma$ values}

Here we move from the variable $\delta$ to the anomalous dimesion $\gamma$ to
get an exact match with the results given in \cite{Alfimov:2023vev}.
Since $\gamma=2\delta -1$, then we have the exact result
\be
 C_{I}(\gamma)=C_{I}(-(\gamma+1))=\frac{1}{(32\pi^2)}\,
 \int_1^{\infty}\,dz\,\left[z^{(\gamma-1)/2}+z^{-1-\gamma/2}\right]\,\frac{1}{\sqrt{1+z}}\,
\ln \frac{\sqrt{1+z}+1}{\sqrt{1+z}-1}\,
  \label{CIdelta.g}
\ee
as well as the approximation for small $\gamma$ values (it can be
obtained from (\ref{CIdelta.O3}) by formally replacing
$2\delta \to -\gamma$)
\be
C_I(\gamma)=
%C_{I}(1/2-\delta)=
\frac{1}{64\pi^2} \bigl[C_0-\gamma\, C_1 + \gamma^2\, C_2 + O(\gamma^3)\bigr]\,,
\label{CIdelta.O3g}
\ee
where $C_i$ $(i=0,1,2)$ are given in Eq. (\ref{Ci}).

Introducing $\eta=\zeta/(1024\pi^2)$ as in Ref. \cite{Alfimov:2023vev}, we have
\be
\eta(\gamma)\equiv \frac{\zeta(\gamma)}{1024\pi^2}=- \frac{\pi^2\gamma(1+\gamma)}{\bigl[C_0-\gamma\, C_1 + \gamma^2\, C_2\,
    %+ O(\delta^3)
    \bigr]}\,.
\label{nu}
\ee

So, we see that
\be
\frac{1}{\eta}=-\frac{1}{\pi^2\gamma(1+\gamma)}\, \bigl[C_0-\gamma\, C_1 + \gamma^2\, C_2\,+ O(\gamma^3)   \bigr]= -\frac{1}{\gamma(1+\gamma)}\, \bigl[1-\gamma\,
  \hat{C}_1 + \gamma^2\, \hat{C}_2\,+ O(\gamma^3)   \bigr]\,,
\label{1nu}
\ee
where $\hat{C}_i=C_i/C_0$ and
\be
\hat{C}_1=l_2-\frac{21\zeta_3}{2\pi^2},~~\hat{C}_2=\frac{\pi^2}{40}+\,\frac{l_2^4}{2\pi^2}+12\frac{{\rm Li}_4(1/2)}{\pi^2}\,.
\label{Ci}
\ee

Eq. (\ref{1nu}) can be rewritten as
\be
\frac{1}{\eta}=-\frac{1}{\gamma}\, \bigl[1-\gamma\,
  \overline{C}_1 + \gamma^2\,\overline{C}_2\,+ O(\gamma^3)   \bigr]\,,
\label{1nu.1}
\ee
where
\be
\overline{C}_1=1+ \hat{C}_1,~~ \overline{C}_2=1+ \hat{C}_1+\hat{C}_2\,.
\label{1nu}
\ee

Inverting Eq. (\ref{1nu.1}), we have
\be
\gamma=-\eta-\overline{C}_1\,\eta^2-\bigl(\overline{C}_2+\overline{C}_1^2\bigr)\,\eta^2 + O(\eta^4)\,,
\label{gamma_nu}
\ee
which is exatly coincides with obtained in \cite{Alfimov:2023vev}.

\subsection{A vicinity of $\gamma_c =-1/2$}

By studying the $\gamma$ behavior in the vicinity of  $\gamma_c=-1/2$,
authors of \cite{Alfimov:2023vev} found
%We see that
\be
\gamma \approx -\frac{1}{2} \pm iC \sqrt{\zeta - \zeta_{c}}\,,
%~\to
%\gamma+ \frac{1}{2}=\pm iC \sqrt{\zeta - \zeta_{c}}\nonumber \\
%\to \left(\gamma+ \frac{1}{2}\right)^2=-C^2\,(\zeta - \zeta_{c})
 \label{gamma_zeta} 
\ee
where $C$ is some constant. We will carefully study the behavior and
find the $C$ value.

From (\ref{gamma_zeta}) we see that
\be
%\gamma \approx -\frac{1}{2} \pm iC \sqrt{\zeta - \zeta_{c}}\,,
%~\to
\gamma+ \frac{1}{2}=\pm iC \sqrt{\zeta - \zeta_{c}}~~~\mbox{and, thus,}~~~
%\nonumber \\
%\to
\left(\gamma+ \frac{1}{2}\right)^2=-C^2\,(\zeta - \zeta_{c})
 \label{gamma_zeta1} 
\ee

With another side, $\gamma=2\delta-1$ and so
\be
%\gamma=2\delta-1 \to
\gamma+ \frac{1}{2}=2\delta - \frac{1}{2}=2\ep
 \label{gamma_delta} 
\ee

Thus, we obtain that
\be
4\ep^2=-C^2\,(\zeta - \zeta_{c})~~~\mbox{, or}~~~
%\to
\zeta - \zeta_{c}= -\frac{4}{C^2}\,\ep^2
 \label{gamma_delta} 
\ee

According to Eq. (\ref{xi_delta_ap}), we have
%With another side:
\be
\zeta(\delta)
%\frac{\delta(1-2\delta)}{8C_I(\delta)}=\frac{1-16\ep^2}{64C_I(1/4+\ep)}
=\frac{1}{64C_I(1/4)}\,
\left(1-\ep^2\left[16+1/2\,L_I(1/4)\right]+O(\ep^4)\right)\,,
 \label{zeta_ep} 
\ee
where
\be
L_I(1/4)=C^{(1)}_I(1/4)/C_I(1/4)\approx 4.2198,~~\mbox{since}~~C^{(1)}_I(1/4)\approx 0.057386
\label{LI14} 
\ee
and $C_I(1/4)$ is given in Eq. (\ref{CI14o}).
Comparing Eqs. (\ref{gamma_delta}) and (\ref{zeta_ep}), we see that
%So, we have
\be
\zeta_{c}=\frac{1}{64C_I(1/4)}\,,~~C^2=\frac{16C_I(1/4)}{1+1/32\,L_I(1/4)}=\frac{16C^2_I(1/4)}{C_I(1/4)+1/32\,C^{(1)}_I(1/4)}\,
\label{zeta_cC2} 
\ee
and, thus,
\be
C=\frac{4C_I(1/4)}{\sqrt{C_I(1/4)+1/32\,C^{(1)}_I(1/4)}}\approx 0.43845\,.
\label{CCI} 
\ee

Since $L_I(1/4)\approx 4.2198 << 32$, then the following approximation
can be taken
\be
C\approx 4\sqrt{C_I(1/4)}=\frac{1}{2\sqrt{\zeta_{c}}}\approx 0.46646\,.
\label{C_zeta_c} 
\ee

\section{Summary}

We studied the shortest single-trace operator
in the fishnet limit of the twisted
  ABJM model in 3D and found its anomalous dimension in the simple
  form (\ref{xi_delta_ge}).

  Indeed, using the Fourier transform, we were able to present
  the results for the eigenvalue $h(\nu)$ of the spectral equation
  for the operator in the form containing ${}_3F_{2}$-hypergeometric
  function with argument 1.
We also calculated the results around the critical point
$\delta_c=1/4$, as well as for small values of the corresponding
anomalous dimension $\gamma$.
We have discovered the property of uniform transcendentality in the
coefficients of the $\gamma$-expansion of the ratio
$h(\gamma)/[\gamma(1+\gamma)]$.

Numerically we have an exact agreement
with the results obtained in \cite{Alfimov:2023vev}.\\

\appendix

\def\theequation{A\arabic{equation}}
\setcounter{equation}{0}
\section{Evaluating $S(\delta)$}

Now we consider $S(\delta)$:
\be
S(\delta)=
\sum_{n=0}^{\infty} \,
  \frac{\Gamma(n+1)\Gamma(\delta)}{\Gamma(n+\delta+1)}\,
  \int_0^1\,dx\,x^{n+\delta-1/2}=\int_0^1\,dx\,\frac{x^{\delta-1/2}}{\delta} \, {}_2F_1\bigl(1,1;1+\delta;x\bigr)\,,
   \label{Sdelta}
\ee
where $ {}_2F_1$ is Gauss hypegeometric function.

Using its property:
\be
{}_2F_1\bigl(a,b;c;x\bigr)=(1-x)^{-a}\,{}_2F_1\bigl(a,c-b;c;x/(x-1)\bigr)\,,
   \label{F21}
\ee
we have for $S(\delta)$:
\be
S(\delta)=\int_0^1\,dx\,\frac{x^{\delta-1/2}}{\delta(1-x)} \, {}_2F_1\bigl(1,\delta;1+\delta;x/(x-1)\bigr)
=\int_0^1\,dx\,\frac{x^{\delta-1/2}}{(1-x)} \, \sum_{n=0}^{\infty} \,\frac{1}{(n+\delta)}\, \left(\frac{x}{x-1}\right)^n\,,
   \label{Sdelta.1}
\ee
where
\be
\sum_{n=0}^{\infty} \,\frac{1}{(n+\delta)}\, \left(\frac{x}{x-1}\right)^n=\sum_{n=0}^{\infty} \,\left(\frac{x}{x-1}\right)^n\,\int_0^1\,dt\, t^{n+\delta-1}
=\int_0^1\,dt\, t^{\delta-1}\,\frac{1-x}{1-x+tx} \,.
   \label{Sum}
\ee

Introducing the new variable $z=xt$, we have for the r.h.s.
\be
\int_0^x\,dz\, \frac{z^{\delta-1}}{x^{\delta}}\,\frac{1-x}{1-x+z} \,
   \label{Sum.1}
\ee
and, thus, for  $S(\delta)$ (with $x=p^2$)
\be
S(\delta)=\int_0^1\,dz\, z^{\delta-1}\, \int_z^1\,\frac{dx}{x^{1/2}}\,\frac{1}{1-x+z} \,=2\, \int_0^1\,dz\, z^{\delta-1}\, \int_{\sqrt{z}}^1\,
\frac{dp}{1-p^2+z}\,, 
   \label{Sdelta.2}
\ee
where now the inner integral does not depend on $\delta$. It can be calculated as
\be
\int_{\sqrt{z}}^1\, \frac{dp}{1-p^2+z}\, = \frac{1}{2\sqrt{1+z}}\, \ln \frac{\sqrt{1+z}+p}{\sqrt{1+z}-p}|^1_{\sqrt{z}}=
\frac{1}{2\sqrt{1+z}}\, \ln \frac{(\sqrt{1+z}+1)(\sqrt{1+z}-\sqrt{z})}{(\sqrt{1+z}-1)(\sqrt{1+z}+\sqrt{z})}
   \label{Intz}
\ee
and, thus,
\be
S(\delta)= \int_0^1\,dz\,\frac{z^{\delta-1}}{\sqrt{1+z}}\, \ln \frac{(\sqrt{1+z}+1)(\sqrt{1+z}-\sqrt{z})}{(\sqrt{1+z}-1)(\sqrt{1+z}+ \sqrt{z})}\,.
   \label{Sdelta.3}
\ee

We would like to note that using the replacement $z\to 1/z$ we can rewrite
\be
\int_0^1\,dz\,\frac{z^{\delta-1}}{\sqrt{1+z}}\, \ln \frac{\sqrt{1+z}-\sqrt{z}}{\sqrt{1+z}+ \sqrt{z}}=
\int_1^{\infty}\,dz\,\frac{z^{-1/2-\delta}}{\sqrt{1+z}}\, \ln \frac{\sqrt{1+z}-1}{\sqrt{1+z}+1}
\label{Intz.1}
\ee
and, thus,
\be
S(\delta)= \left[\int_0^1\,dz\,z^{\delta-1}-\int_1^{\infty}\,dz\,z^{-1/2-\delta}\right]\,\frac{1}{\sqrt{1+z}}\,
\ln \frac{\sqrt{1+z}+1}{\sqrt{1+z}-1}\,.
   \label{Sdelta.4}
   \ee

   Now we calculate the following integral:
\be
\int_0^{\infty}\,dz\,\frac{z^{\delta-1}}{\sqrt{1+z}}\, \ln \frac{\sqrt{1+z}+1}{\sqrt{1+z}-1}
=\frac{\sqrt{\pi}\Gamma(1-\delta)\Gamma^2(\delta)}{\Gamma(1/2+\delta)}\,
   \label{Intx.2}
\ee
which is exacly equal to the first term in the r.h.s. of Eq. (\ref{CIdelta}). 
Thus, we have Eq. (\ref{CIdelta.1}) for $C_I(\delta)$.

\def\theequation{B\arabic{equation}}
\setcounter{equation}{0}
%\renewcommand\theequation{A.\arabic{equation}}
%\section{ Basic formulas for massless diagrams}
\section{Evaluating $\overline{S}(\delta)$ }

Now we consider $\overline{S}(\delta)$ in the form
\be
\overline{S}(\delta)= 
\sum_{n=0}^{\infty} \,
  \frac{\Gamma(n+1)\Gamma(1/2-\delta)}{\Gamma(n+3/2-\delta)}\,
%  \int_0^1\,dx\,x^{n+\delta-1/2}=\int_0^1\,dx\,\frac{x^{\delta-1/2}}{\delta} \, {}_2F_1\bigl(1,1;1+\delta;x\bigr)\,,
\sum_{s=0}^{\infty}\, \frac{\delta^s}{(n+1)^{s+1}}\,.
  \label{oSdelta.C}
\ee

Using the following property:
\be
\frac{1}{(n+1)^{s+1}}=\frac{(-1)^s}{s!}\, \int_0^1\,dx\,x^{n}\, \ln^sx\,,
  \label{Int}
\ee
we have
\be
\overline{S}(\delta)= \sum_{s=0}^{\infty}\, \frac{(-\delta)^s}{s!} \, \int_0^1\,dx\, \ln^sx\,
\sum_{n=0}^{\infty} \
  \frac{\Gamma(n+1)\Gamma(1/2-\delta)}{\Gamma(n+3/2-\delta)}\,x^n\,,
%  \int_0^1\,dx\,x^{n+\delta-1/2}=\int_0^1\,dx\,\frac{x^{\delta-1/2}}{\delta} \, {}_2F_1\bigl(1,1;1+\delta;x\bigr)\,,
%\sum_{s=0}^{\infty}\, \frac{\delta^s}{(n+1)^{s+1}}\,.
  \label{oSdelta.C1}
  \ee
  where
  \bea
 && \sum_{n=0}^{\infty} \,
  \frac{\Gamma(n+1)\Gamma(1/2-\delta)}{\Gamma(n+3/2-\delta)}\,x^n= \frac{1}{1/2-\delta} \, {}_2F_1\bigl(1,1;3/2-\delta;x\bigr)\nonumber\\
&&=  \frac{(1-x)^{-1}}{1/2-\delta} \, {}_2F_1\bigl(1,1/2-\delta;3/2-\delta;x\bigr)
=  \frac{1}{(1-x)}\, \sum_{n=0}^{\infty} \,
\frac{1}{n+1/2-\delta}\,\left(\frac{x}{x-1}\right)^n \nonumber\\
&&=   \frac{1}{(1-x)}\, \sum_{n=0}^{\infty} \,
  \int_0^1\,dt\,t^{n-1/2-\delta}\,\left(\frac{x}{x-1}\right)^n=   \int_0^1\,dt\,\frac{t^{-1/2-\delta}}{1-x+tx}
 \label{Sum.C}
  \eea

  Here $ {}_2F_1$ is Gauss hypegeometric function and we used the property (\ref{F21}). In a sence, the results are obtained in a close analogy results
  done in Appendix A.

So,  we have for $\overline{S}(\delta)$:
\be
\overline{S}(\delta)= \sum_{s=0}^{\infty}\, \frac{(-\delta)^s}{s!} \, \int_0^1\,dx\,ln^sx\,\int_0^1\,dt\,\frac{t^{-1/2-\delta}}{1-x+tx}\,.
   \label{oSdelta.C2}
\ee

It is convenient to consider firstly the case with $s=0$. We have
\be
 \int_0^1\,dx\,\int_0^1\,dt\,\frac{t^{-1/2-\delta}}{1-x+tx}
   \label{s0}
\ee

After integrationg on $x$, we obtain (with a small parameter $\overline{\ep}$)
\be
-\int_0^1\,dt\,\frac{t^{-1/2-\delta}\ln t}{(1-t)^{1-\overline{\ep}}}=-\frac{d}{d\delta}\, \int_0^1\,dt\,\frac{t^{-1/2-\delta}}{(1-t)^{1-\overline{\ep}}}
= \frac{\Gamma(1/2-\delta)\Gamma(\overline{\ep})}{\Gamma(1/2-\delta-\overline{\ep})}\, \bigl(\Psi(1/2-\delta-\overline{\ep})-\Psi(1/2-\delta)\bigr)
\label{s0.1}
\ee

In the limit $\overline{\ep} \to 0$, we have
\be
\int_0^1\,dx\,\int_0^1\,dt\,\frac{t^{-1/2-\delta}}{1-x+tx}=\Psi^{'}(1/2-\delta)=\sum_{m=0}^{\infty}\, \frac{(-\delta)^m}{m!}\,\Psi^{m+1}(1/2)= 3\zeta_2+14\zeta_3\delta + ...
\label{s0.2}
\ee

Now we consider the general case.
\be
\int_0^1\,dx\,\frac{\ln^s x}{1-x+tx}= \frac{s!}{(-1)^s}\, \frac{{\rm Li}_{s+1}(1-t)}{1-t}\,,
\label{gs}
\ee
where ${\rm Li}_{s+1}$ is the Polylogarithm.

So, we have (with $t=p^2$)
\bea
\overline{S}(\delta)&=& \sum_{s=0}^{\infty}\, \delta^s \, \int_0^1\,dt\,\frac{t^{-1/2-\delta}}{1-t}\,{\rm Li}_{s+1}(1-t)
= 2 \sum_{s=0}^{\infty}\, \delta^s \, \int_0^1\,dp\,\frac{p^{-2\delta}}{1-p^2}\,{\rm Li}_{s+1}(1-p^2)\nonumber \\
&=& 2 \sum_{s=0}^{\infty}\, \delta^s \, \int_0^1\,\frac{dp}{1-p^2}\,\sum_{m=0}^{\infty}\,\frac{(-2)^m}{m!}\,\ln^mp\,{\rm Li}_{s-m+1}(1-p^2)
\label{oSdelta.C3}
\eea

%\end{document}

\end{document}